\documentclass{PoS}

\title{A GEM-based Time Projection Chamber for the AMADEUS experiment}

\ShortTitle{TPC-GEM for AMADEUS experiment}

\author{\speaker{M.~Poli Lener}
        Laboratori Nazionali di Frascati - INFN, Frascati, Italy\\
        E-mail: \email{marco.polilener@lnf.infn.it}}

\author{M.~Bazzi, G.~Corradi, C.~Curceanu, A.~D'Uffizi, C.~Paglia, E.~Sbardella,  A.~Scordo, D.~Tagnani\\
        Laboratori Nazionali di Frascati - INFN, Frascati, Italy}
\author{A.~Romero Vidal,\\
        Laboratori Nazionali di Frascati - INFN, Frascati, Italy\\
        Universidad de Santiago de Compostela, Santiago de Compostela, Spain}
\author{J.~Zmeskal\\
        Stefan Meyer Institut fur s\"{u}batomare Physik, Vienna, Austria}

\abstract{
In this paper we present the R\&D activity on a new GEM-based TPC prototype
for AMADEUS, a new experimental proposal at the DA$\Phi$NE 
$\Phi$-factory at the Laboratori Nazionali di Frascati (INFN), 
aiming to perform measurements of the low-energy negative kaons interactions 
in nuclei.
Such innovative detector will equip the inner part of the experiment
in order to perfom a better reconstruction of the primary vertex and the
 secondary particles tracking.\\ 
A 10x10 cm$^2$ prototype with a drift gap up to 15 cm was realized
and succesfully tested at the $\pi$M1 beam facility of the Paul Scherrer 
Institut (PSI) with low momentum hadrons.
The measurements of the detector efficiency and spatial resolution have been
performed. The results as a function of the gas gain, drift field, 
front-end electronic threshold and particle momentum are reported and discussed.}

\FullConference{International Winter Meeting on Nuclear Physics,\\
		21-25 January 2013\\
		Bormio, Italy }

\begin{document}

\section{Introduction}
\label{sec:intro}
\begin{figure}[ht!]
  \centering
    \includegraphics[width=6.cm]{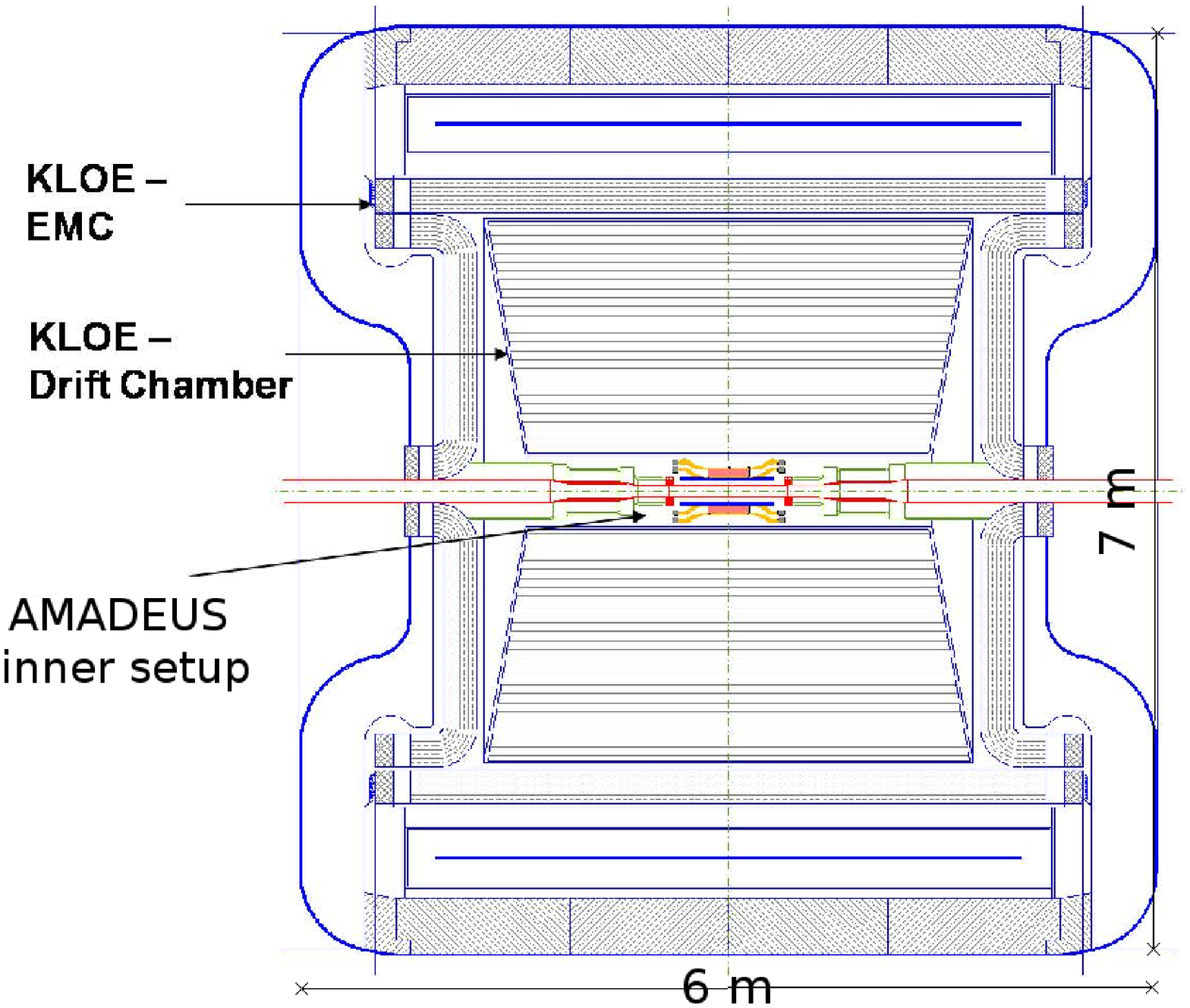}%
    \qquad
    \includegraphics[width=8cm]{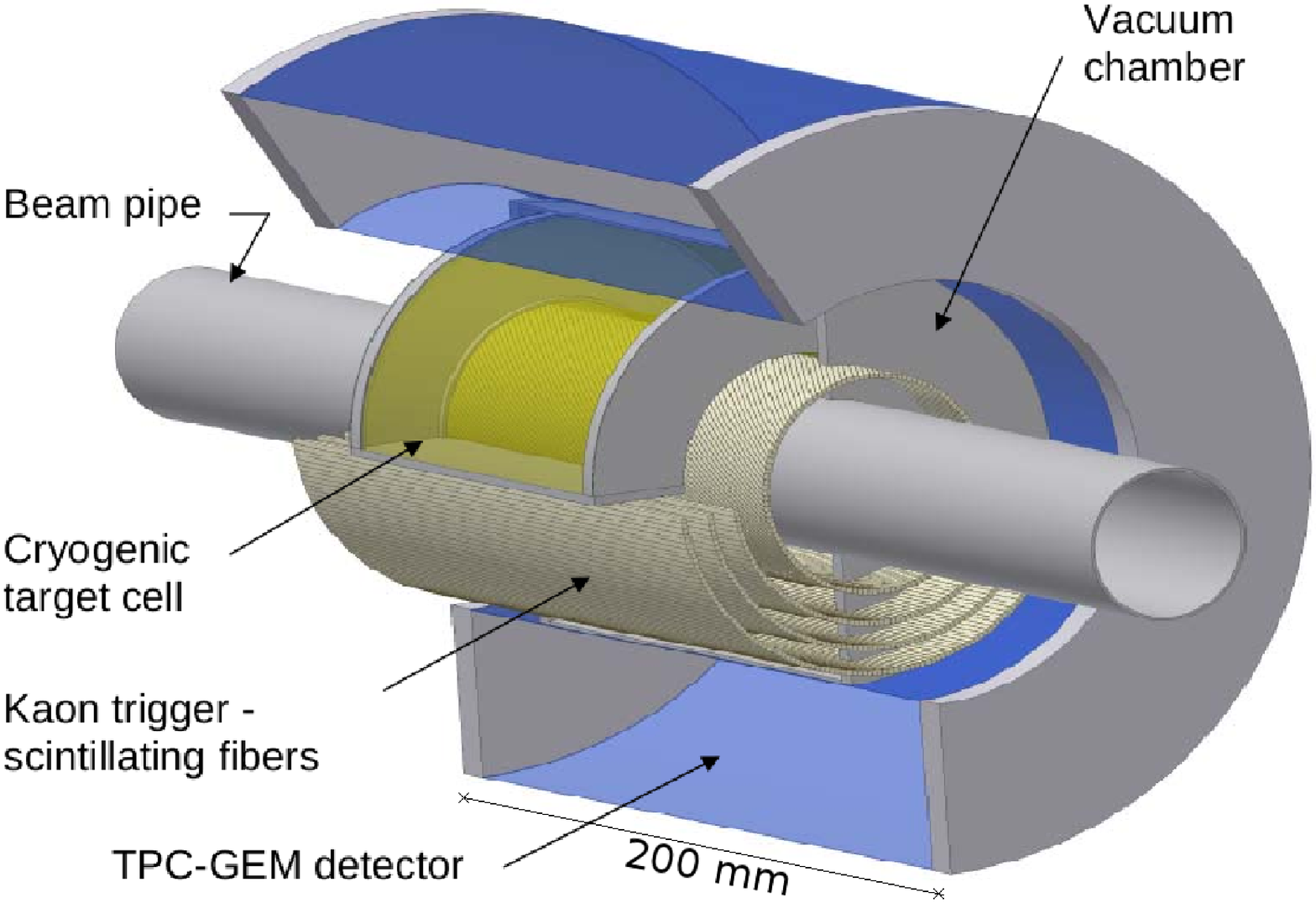}%
     \caption{Cross-section of the KLOE detector including the 
      AMADEUS inner setup inside the Drift Chamber (left) and 
              the inner setup of AMADEUS. From the beam pipe to
        the outer region: the Kaonic Trigger, the Gas Target and the GEM based-TPC.}
      \label{fig:amadeus}
\end{figure}

An important, yet unsolved problem, in hadron physics is how 
the hadron masses and interactions change in the nuclear medium. 
This topic could be investigated by means of ``in-medium hadron-mass 
spectroscopy'', producing bound states of a hadron by which to 
deduce the hadron-nucleus potential and the in-medium hadron masses. 
The AMADEUS (Antikaon Matter At DA$\Phi$NE Experiments with Unraveling 
Spectroscopy) experiment~\cite{AMADEUS},\cite{AMADEUS1} will study 
the low energy interactions of kaons with nucleons and nuclei.
The AMADEUS setup will be implemented inside the KLOE~\cite{KLOE} Drift Chamber (DC), 
in the free space between the beam pipe and the DC entrance wall. Three main 
components of the experimental setup are under development: a high density 
cryogenic gaseous target, a trigger system~\cite{scordo}, and an inner 
tracker, namely a Time Projection Chamber equipped with Gas Electron 
Multiplier~\cite{sauli}, to be positioned inside the KLOE DC, which will
perform a better reconstruction of the primary vertex and the
 secondary particles tracking.\\
A representation of the dedicated AMADEUS setup surrounding the beam pipe 
within KLOE detector is given in Figure~\ref{fig:amadeus}.\\

The GEM-based TPC (TPG) will be 20 cm long with an inner diameter of 8 cm and
 an outer one of 40 cm. The main requirements for the TPG can be 
summarized as: 
\begin{itemize}
  \item a spatial resolution better than 200 $\mu$m in X-Y and 
        300 $\mu$m in Z;
  \item a detector material budget lower than 0.5$\%$ of X$_0$;
  \item a rate capability of $\sim$ 5 kHz/cm$^2$~\cite{MPL};
\end{itemize}
Since most of the above requirements are easily fulfilled by a TPG, 
the R\&D activity at the Laboratori Nazionali di Frascati (INFN) is 
mainly focused on the choice of the gas mixture, in order to achieve 
the highest spatial resolution with a 0.5 T value of magnetic field, 
and on the design of the detector readout.\\
A TPG prototype of 10$\times$10 cm$^2$ with a drift gap up to 15 cm was 
realized and tested both in laboratory 
and at Paul Scherrer Institut (PSI). Detail of the prototype construction
can be found in Ref.~\cite{arxiv}.
The experimental setup of the test beam and the choice of the gas mixtures 
are described in Sec.~\ref{sec:setup}. 
In Sec.~\ref{sec:results} the detector performances obtained at the $\pi$M1 beam 
facility, in terms of detector efficiency and spatial resolution, 
are presented and discussed. The paper ends with conclusions.
\section{Experimental setup}
\label{sec:setup}
The performance of the TPG prototype was studied at the 
$\pi$M1 beam facility of the PSI without magnetic field. The $\pi$M1 beam 
is a quasi-continuous high-intensity secondary beam providing hadrons
with a momentum resolution of $\sim$ 1\%.\\
The measurement of the detector efficiency and spatial resolution 
was performed with a beam rate of $\sim$ 200 Hz.\\
The prototype readout is composed by 4 rows of 32 pads
of $\sim$ 3x3 mm$^2$ each. Each pad  was connected to a front-end board based 
on CARIOCA-GEM chip~\cite{carioca}.
The discriminator threshold on the front-end electronics 
was set to $\sim$ 3.5 fC and $\sim$ 5 fC.\\
During this test beam isobutane-based gas mixtures were used. As shown in
Tab.~\ref{tab:cluster}, such gas mixtures are charactezed by:
\begin{itemize}
 \item a large primary ionisation;
 \item a high drift velocity;
 \item a high Townsend coefficient;
 \item a moderate longitudinal and transversal diffusions;
 \item a very low attachment coefficient;
\end{itemize}

\begin{table}[!ht] 
\centering \scriptsize 
%\tiny   
\begin{tabular}{c | c | c | c | c c c} 
\hline
%\hline $\sigma_{long}$~~~~ \&~~~~ $\sigma_{tran}$ coeff. 
Gas                     &  Average            &  Drift Velocity & Longitudinal \& Transversal Diff. &           & Cluster/cm&    \\
Mixture                 &  Townsend Coeff.    &  at 200 V/cm    & at 200 V/cm                           & 170 MeV/c & 440 MeV/c & MIPs\\
                        & [1/V]              & [$\mu$m/ns]    & [$\mu$m/$\sqrt{cm}$]                  & Pion      & Proton          \\[1.ex] %adds vertical space
%\\[1.ex] %adds vertical space
\hline %\\[1.ex] %adds vertical space
% Entering 1st row 
Ar/C$_4$H$_{10}$ = 80/20 & \bf (25.2$\pm$0.2)$\times$10$^{-3}$ & \bf 29$\pm$2  & 217$\pm$16~~~~~~~~274$\pm$12           &45.2$\pm$2.1 &96.6$\pm$3.5&40.0$\pm$2.0 \\[1.ex] %adds vertical space 

\hline	% inserts single-line 
%\\[1.5ex] %adds vertical space

% Entering 2nd row 
Ar/C$_4$H$_{10}$ = 90/10 & \bf (27.2$\pm$0.3)$\times$10$^{-3}$  & \bf 39$\pm$2   & 282$\pm$7~~~~~~~~359$\pm$18          &37.2$\pm$1.9&79.6$\pm$2.8&32.8$\pm$1.8  \\[1.5ex] %adds vertical space 
\hline	% inserts single-line 
%\\[1.5ex] %adds vertical space

% Entering 3rd row 
Ar/CO$_2$ = 70/30      & \bf(22.0$\pm$0.3)$\times$10$^{-3}$  &4.51$\pm$0.02     & 178$\pm$11~~~~~~~~175$\pm$6              &32.2$\pm$1.8&68.8$\pm$2.6&28.4$\pm$1.6  \\[1.5ex] %adds vertical space 
%\\[1ex] %adds vertical space 
\hline	% inserts single-line 
\end{tabular} 

\caption{Average of Townsend coefficient, drift velocity, diffusion coefficients and primary ionization.
         The values in bold were measured experimentally~\cite{arxiv}, 
         while the normal ones were obtained by Garfield simulation~\cite{garfield}.
         The Ar/CO$_2$=70/30 gas mixture and MIPs are reported for 
         comparison.} % title name of the table
\label{tab:cluster} 
\end{table}

\section{Detector Performances}
\label{sec:results}

\subsection{Efficiency}
\label{sec:efficiency}

\begin{figure}[ht!]
  \centering
%    \includegraphics[width=6.9cm]{effvsdet.eps}%
%    \qquad
    \includegraphics[width=9cm]{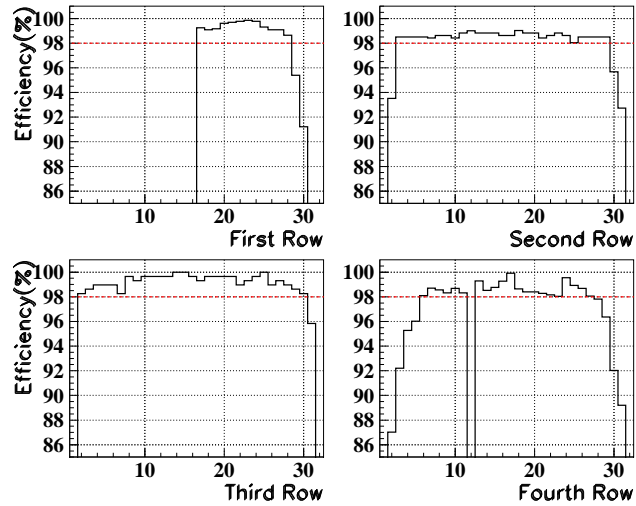}%
     \caption{Single pad efficiency.
% with the relative statistical error.        
      Each pad is $\sim$3$\times$3 mm$^2$.}
      \label{fig:efficiency}
\end{figure}

\begin{figure}[ht!]
  \centering
  \includegraphics[width=9cm]{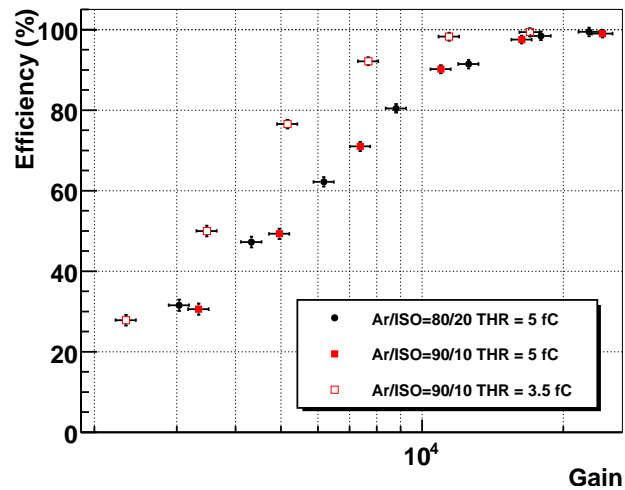}    
    \caption{Detector Efficiency for the Ar/C$_4$H$_{10}$= 80/20 and
     Ar/C$_4$H$_{10}$= 90/10 gas mixtures as a function of the gas gain 
     with a fixed drift field of 150 V/cm and for a 170 MeV/c pion beam.}
    \label{fig:effvsgain}
\end{figure}

The single pad row detector efficiency was evaluated considering 
the fraction of the hits in a single pad row with respect to a selected track.\\
In Fig.~\ref{fig:efficiency} is shown the single pad efficiency for each row. 
It is worth noticing that most of the pads have an efficiency larger 
than 98\% except for the first 16 channels of the first row and 
one channel of the fourth row that were dead. 
A low and/or not full efficiency in the first and last pads 
of each rows and a parabolic behaviour are clearly visible. 
These effects were discussed in detail in Ref.~\cite{arxiv}.\\  
The detector efficiency for the Ar/C$_4$H$_{10}$= 80/20 and 
Ar/C$_4$H$_{10}$= 90/10 gas mixtures as a function of the gas gain with 
150 V/cm drift field and 170 MeV/c pion beam is shown in 
Fig.~\ref{fig:effvsgain}.
As expected, the use of 3.5 fC front-end electronics threshold allows to
reach a full efficiency at lower values of gain 
with respect to the higher threshold measurements.

\begin{figure}[ht!]
  \centering
  \includegraphics[width=9cm]{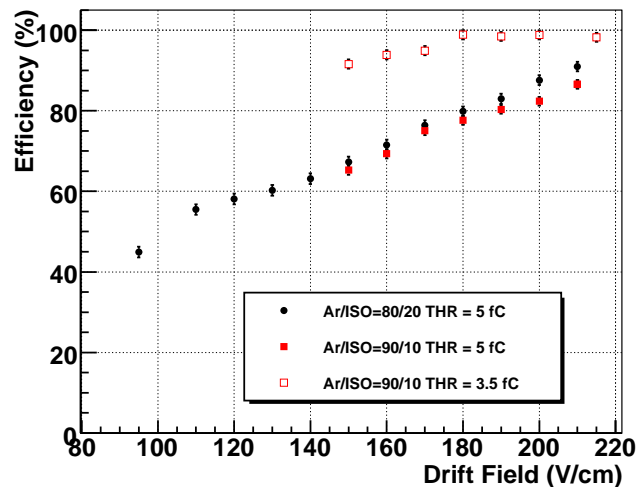}    
    \caption{Detector Efficiency for the Ar/C$_4$H$_{10}$= 80/20 and
     Ar/C$_4$H$_{10}$= 90/10 gas mixtures as a function of the drift field with 
     a fixed gain of $\sim$8$\times$10$^3$ and for a 170 MeV/c pion beam.}
    \label{fig:effvsed}
\end{figure}

The increase of the drift field from 150 V/cm to 210 V/cm with a fixed 
gain of $\sim$ 8$\times$10$^3$  allows to increase the 
detector efficiency from $\sim$ 70\% to $\sim$ 90\% for both gas mixtures at 
5 fC threshold and from  $\sim$ 90\% to a full efficiency for 
Ar/C$_4$H$_{10}$= 90/10 gas mixture at 3.5 fC threshold, as shown 
in Fig.~\ref{fig:effvsed}.
This effect can be explained by a greater collection efficiency 
of the primary electrons in the first GEM: indeed, when the drift 
field increases in the above mentioned range the diffusion of the
primary electrons decreases allowing to reach the GEM holes and to
give rise to the avalanche process.\\
The different level of efficiency between the two curves 
at 5 fC threshold is due to the higher number of primary electrons 
produced in the drift gap in the Ar/C$_4$H$_{10}$= 80/20 
gas mixture with respect to the Ar/C$_4$H$_{10}$= 90/10 one 
(see Tab.~\ref{tab:cluster}).

\begin{figure}[h!]
  \centering
    \includegraphics[width=9cm]{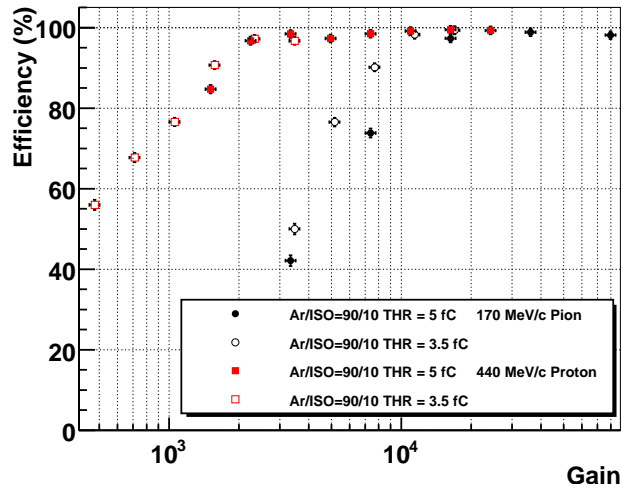}
    \caption{Detector Efficiency for protons of 440 MeV/c and pions
             of 170 MeV/c as a function of the gas gain. The gas mixture is 
             Ar/C$_4$H$_{10}$= 90/10 and the drift field is set to 150 V/cm.}
    \label{fig:effvspar}
\end{figure}

The efficiency for 440 MeV/c proton and 170 MeV/c pion beams for
Ar/C$_4$H$_{10}$= 90/10 gas mixture is shown in Fig.~\ref{fig:effvspar}. 
As expected, protons allow to reach an efficiency plateau at lower gas gain 
with respect to pions due to a higher primary ionisation.   
In addition, the detector efficiency with a proton beam does not seem affected 
by the electronic threshold. This is due to a high primary ionisation of protons 
which allows to produce a signal above the used discriminator thresholds.\\ 
On the contrary, the low threshold value with a pions beam allows to increase
the detector efficiency which reaches an efficiency plateau at lower gas 
gain with respect to the 5 fC measurement.

\subsection{Spatial Resolution}
\begin{figure}[ht!]
  \centering
    \includegraphics[width=9cm]{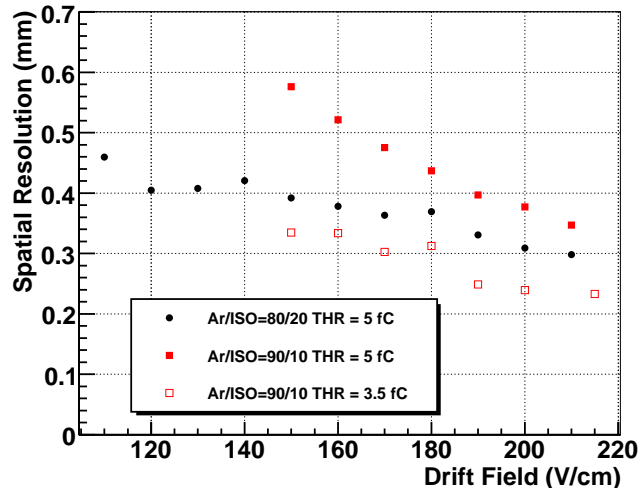}%res1.eps
     \caption{Spatial resolution for 
      the Ar/C$_4$H$_{10}$= 80/20 and Ar/C$_4$H$_{10}$= 90/10 
      gas mixtures as a function of
      the drift field with a fixed gain of $\sim$8$\times$10$^3$ 
      and for a 170 MeV/c pion beam.}
      \label{fig:resvsed}
\end{figure}

The spatial resolution in the drift direction was evaluated by the residuals 
between the hit position in a single pad row and a selected track.\\
%As shown in Fig.~\ref{fig:resvsed} a better spatial resolution 
%is achieved for high drift field and low value of threshold for 170 MeV/c pions. 
Fig.~\ref{fig:resvsed} shows the spatial resolution in the drift direction 
for 170 MeV/c pions as a function of the drift field.
Since the diffusion decreases by increasing the drift field, i.e a greater 
collection efficiency of primary electrons into the first GEM holes,
 a better spatial resolution is achieved for a high drift field in the 
range between 100 and 210 V/cm and a low value of the threshold.\\
Due to a lower diffusion of the Ar/C$_4$H$_{10}$= 80/20 gas mixture with 
respect to Ar/C$_4$H$_{10}$= 90/10 one, a better spatial 
resolution is reached at 5 fC of threshold.

\begin{figure}[ht!]
  \centering
    \includegraphics[width=9cm]{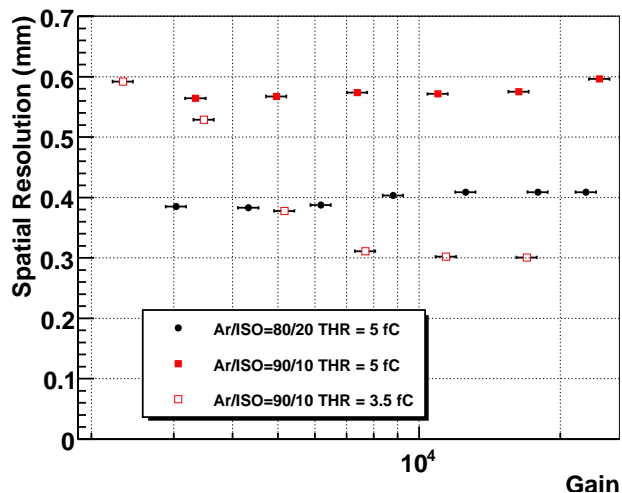}%res.eps
     \caption{Spatial resolution for 
      the Ar/C$_4$H$_{10}$= 80/20 and Ar/C$_4$H$_{10}$= 90/10 
      gas mixtures as a function of gas gain with a fixed drift field of 150 V/cm
      and 170 MeV/c pions.}
      \label{fig:resvsgain}
\end{figure}

The spatial resolution for 170 MeV/c pions as a function of the gas 
gain at fixed drift field is shown in Fig.~\ref{fig:resvsgain}.
For high threshold and for both gas mixtures the spatial resolution seems to be not 
affected by the detector gain, while this does not occur with a lower threshold
value. 
A possible explanation is that the spatial resolution increases 
as the detector gain increases, which allow to increase the collection efficiency  of primary ionisation
into the first GEM holes~\cite{cardini}, and 
until the signal is above the discrimination threshold. 
When this happens a better spatial resolution 
is achieved while it reaches a plateau when the signal is comparable 
with the electronic threshold.\\
Such explanation is confirmed by the Ar/C$_4$H$_{10}$= 90/10
gas mixture measurements, in which the spatial resolution 
reaches the same level at very 
low gain ($<$3$\times$10$^3$) and regardless of the used threshold.\\

\begin{figure}[h!]
  \centering
    \includegraphics[width=9cm]{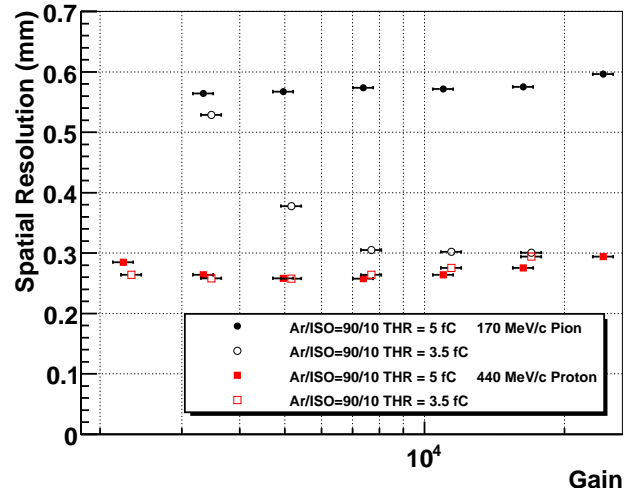}
    \caption{Spatial resolution for proton of 440 MeV/c and pion of 170 MeV/c 
             as a function of the gas gain. The gas mixture is 
             Ar/C$_4$H$_{10}$= 90/10 and the drift field is set to 150 V/cm.}
    \label{fig:resvspar}
\end{figure}

The effect of the ionazing particle on the spatial resolution was 
evaluated comparing 440 MeV/c proton beam with 170 MeV/c pion one, as shown in 
Fig.~\ref{fig:resvspar}. As explained above, the spatial resolution 
seems not so sensible to a gain change with 5 fC threshold value, 
regardless of the ionazing particle. Moreover, the larger 
ionization produced by protons with respect to pions allows to reach a better 
spatial resolution of about a factor 2.\\
With a 3.5 fC of threshold, the spatial resolution for pions reaches 
at high gain the same level obtained with protons. This effect suggest that 
the spatial resolution reaches a limit value for the 
Ar/C$_4$H$_{10}$= 90/10 gas mixture.

\section{Conclusion}

The R\&D activity on a GEM-based TPC for the inner region of
the AMADEUS setup has started at the Laboratori Nazionali di Frascati (INFN). 
A TPG prototype with a drift gap up to 15 cm has been successfully 
produced and tested at the $\pi$M1 beam facility of the Paul 
Scherrer Institut with low momentum pion and proton beams.\\
The measurement of the detector performances, in terms of 
efficiency and spatial resolution as a function of the gas gain, drift field, 
front-end electronics threshold and particle momentum, 
has been reported and discussed in detail.
A detection efficiency of 99\% and a spatial resolution of 
240 $\mu$m have been achieved, compatible with the foreseen values for AMADEUS.

\end{document}